\documentclass[aps,prc,twocolumn, double-spaced, showpacs, superscriptaddress]{revtex4}

\usepackage{graphicx,rotating}
\usepackage[latin1]{inputenc}
\usepackage[T1]{fontenc}

\begin{document}

\title{High-precision mass measurements of nickel, copper, and gallium isotopes
and the purported shell closure at \textit{N}=40}

\author{C. Gu\'enaut}
\altaffiliation[Corresponding author; Present address: ]{NSCL,
Michigan State University, East Lansing, MI 48824, USA}
\email[E-mail address: ]{guenaut@nscl.msu.edu}
\affiliation{CSNSM-IN2P3-CNRS, 91405 Orsay-Campus, France}

\author{G.~Audi}
\affiliation{CSNSM-IN2P3-CNRS, 91405 Orsay-Campus, France}

\author{D.~Beck}
\affiliation{GSI, Planckstra\ss e 1, 64291 Darmstadt, Germany}

\author{K.~Blaum}
\affiliation{GSI, Planckstra\ss e 1, 64291 Darmstadt, Germany}
\affiliation{Johannes Gutenberg-Universität, Institut für Physik,
55099 Mainz, Germany}

\author{G.~Bollen}
\affiliation{NSCL, Michigan State University, East Lansing, MI
48824, USA}

\author{P.~Delahaye}
\affiliation{CERN, Physics Department, 1211 Gen\`eve 23,
Switzerland}

\author{F.~Herfurth}
\affiliation{GSI, Planckstra\ss e 1, 64291 Darmstadt, Germany}

\author{A.~Kellerbauer\footnote{Present address: Max Planck Institut für
Kernphysik, Postfach 103980, 69029 Heidelberg, Germany}}
\affiliation{CERN, Physics Department, 1211 Gen\`eve 23,
Switzerland}

\author{H.-J.~Kluge}
\affiliation{GSI, Planckstra\ss e 1, 64291 Darmstadt, Germany}
\affiliation{Physikalisches Institut, Universität Heidelberg,
69120 Heidelberg, Germany}

\author{J.~Libert}
\affiliation{Institut de Physique Nucléaire, IN2P3-CNRS, 91406
Orsay-Campus, France}

\author{D.~Lunney}
\affiliation{CSNSM-IN2P3-CNRS, 91405 Orsay-Campus, France}

\author{S.~Schwarz}
\affiliation{NSCL, Michigan State University, East Lansing, MI
48824, USA}

\author{L.~Schweikhard}
\affiliation{Institut für Physik, Ernst-Moritz-Arndt-Universität,
17487 Greifswald, Germany}

\author{C.~Yazidjian}
\affiliation{GSI, Planckstra\ss e 1, 64291 Darmstadt, Germany}

\date{\today}

\begin{abstract}
High-precision mass measurements of more than thirty neutron-rich
nuclides around the \textit{Z}=28 closed proton shell were
performed with the triple-trap mass spectrometer ISOLTRAP at
ISOLDE/CERN to address the question of a possible neutron shell
closure at \textit{N}=40. The results, for $^{57,60,64-69}$Ni
($Z=28$), $^{65-74,76}$Cu ($Z=29$), and $^{63-65,68-78}$Ga
($Z=31$), have a relative uncertainty of the order of $10^{-8}$.
In particular, the masses of  $^{72-74,76}$Cu have been measured
for the first time. We analyse the resulting mass surface for
signs of magicity, comparing the behavior of \textit{N}=40 to that
of known magic numbers and to mid-shell behavior. Contrary to
nuclear spectroscopy studies, no indications of a shell or
sub-shell closure are found for \textit{N}=40.
\end{abstract}

% insert suggested PACS numbers in braces on next line
% insert suggested keywords - APS authors don't need to do this

%\maketitle must follow title, authors, abstract, \pacs, and \keywords
\maketitle

% body of paper here - Use proper section commands
% References should be done using the \cite, \ref, and \label commands

\section{Introduction}
A striking parallel between the atomic and nuclear systems is the
occurrence of closed shells. The behavior of the atomic system is
largely governed by what can be considered as an infinitely
massive and point-like nucleus. Describing nuclear behavior,
however, is a particularly difficult task given its composition of
neutrons and protons, similar in mass yet different in charge. The
nucleon interaction is so complicated that ground-state properties
are not globally predicted with particularly good precision.  A
property crucial to the understanding of the nuclear system is the
behavior of its shell structure as a function of the varying
composition of protons and neutrons. The fact that shell structure
seems to be modified in systems where the number of neutrons $N$
and the number of protons $Z$ are unbalanced (\textit{i.e}. far
from the equilibrium region of stable nuclides) is one of the key
questions of today's nuclear physics research.

Over the last 20 years, magic numbers have been found to vanish in
certain region of the chart of nuclides, the first one being
\textit{N}\,=\,20 for sodium\,\cite{thi75} and later,
magnesium\,\cite{Gui84}. More recently, \textit{N}\,=\,8
\cite{Oza00,Shi03} and \textit{N}\,=\,28\,\cite{Sar00,Min06} have
also disappeared. Conversely, ``new'' magic numbers such as
\textit{N}\,=\,16 \cite{Oza00} and \textit{N}\,=\,32
\cite{Pri01,Man03a,Man03b} have also been found. One case of
particular interest is that of $N=40$ because of the unexpected
events that have transpired since the first studies in 1982.  At
that time, Bernas \textit{et al.}\,\cite{Ber82} showed that the
first excited state of $^{68}_{28}$Ni$_{40}$ was $0^+$,
establishing a new case of $2^+$ and $0^+$ inversion. This was
compared to the case of $^{40}_{20}$Ca$_{20}$, a doubly-magic
nuclide\,\cite{Eks92} where such an inversion was known.
Consequently, Bernas \textit{et al.} concluded $^{68}$Ni to be
doubly-magic.

In 1995, Broda\,\textit{et al.}\,\cite{Bro95} published a
comprehensive summary of spectroscopy work since 1982 and
elaborated the excited spectrum of $^{68}$Ni, finding the first
excited state to be $0^+$ (as Bernas \textit{et
al.}\,\cite{Ber82}), $2^+$ as the second excited state and a $5^-$
isomeric state. As this is the same situation for the $^{80}$Zr
excited states, they concluded that $^{68}$Ni was spherical,
implying a significant sub-shell closure at $N=40$. Shell-model
predictions of isomeric states near magic nuclides motivated the
experimental investigations of Grzywacz \textit{et
al.}\,\cite{Grz98} in 1998. They discovered many isomeric states
in the vicinity of $^{68}$Ni, further strengthening the case for
its doubly-magic character.  In 1999, $\beta$-decay studies were
carried out by Hannawald \textit{et al.}\,\cite{Han99}, who found
long half-lives for the neighboring isotones (copper, manganese)
at $N=40$ indicating an increase in collectivity. However,
$\beta$-decay studies by Mueller \textit{et al.}\,\cite{Mue99} the
same year showed that the stabilizing effect of $N=40$ disappeared
when moving away from $^{68}$Ni.

The powerful tool of Coulomb excitation was brought to bear on
$^{68}$Ni in 2002 when Sorlin \textit{et al.}\,\cite{Sor02}
measured the $B(E2)$ value (which is the probability of transition
between the ground state $0^+$ and the excited state $2^+$).
$B(E2)$ is expected to be small for magic nuclides which are
difficult to excite, and to be large for deformed nuclides. The
measured $B(E2)$ value was unexpectedly small, reinforcing the
magic nature of $^{68}$Ni. Sorlin\,\textit{et al.} attributed the
lack of corroborating evidence from the mass surface to an erosion
of the $N=40$ sub-shell, erosion confirmed by recent
measurements\,\cite{Per06,Ste06}. However, a concerted theoretical
effort published by Langanke \textit{et al.}\,\cite{Lan03} argued
against the doubly-magic nature of $^{68}$Ni, noting that the
``missing'' $B(E2)$ strength lies at much higher energy (>4\,MeV).

According to Bohr and Mottelson\,\cite{Boh69}: ``In terms of the
expansion of the total binding energy, the shell structure appears
as a small correction compared to the surface energy... Despite
the smallness of these effects on the scale of the total nuclear
energy, they are of decisive importance for the structure of the
low-energy nuclear spectra...'' In the light of these conflicting
experimental and theoretical signatures as well as the relatively
large uncertainty on the binding energies in this interesting
region, high-precision mass measurements were carried out with the
mass spectrometer ISOLTRAP in an attempt to bring some
clarification to this situation. Time-of-flight mass measurements
had been performed in 1994\,\cite{Sei94} but although they gave no
indication that $N=40$ was magic, the precision was insufficient
to be conclusive. The most accurate mass measurements today are
performed in Penning traps\,\cite{Bla06,Sch06} and ISOLTRAP at
CERN has pioneered the application to radioactive
nuclides\,\cite{Bol87,Sto90}. The experimental setup of ISOLTRAP
is presented in section\,\ref{setup}, and the measurements in the
region of $N=40$ and their evaluation are described in
section\,\ref{nucli}. A comparison to mass models follows in
section \,\ref{Models} and the question of $N=40$ is discussed in
the light of the new results in the last section.

\section{The ISOLTRAP setup}\label{setup}
\subsection{Experimental setup}

\begin{figure*}
\begin{center}
\includegraphics[width=16cm]{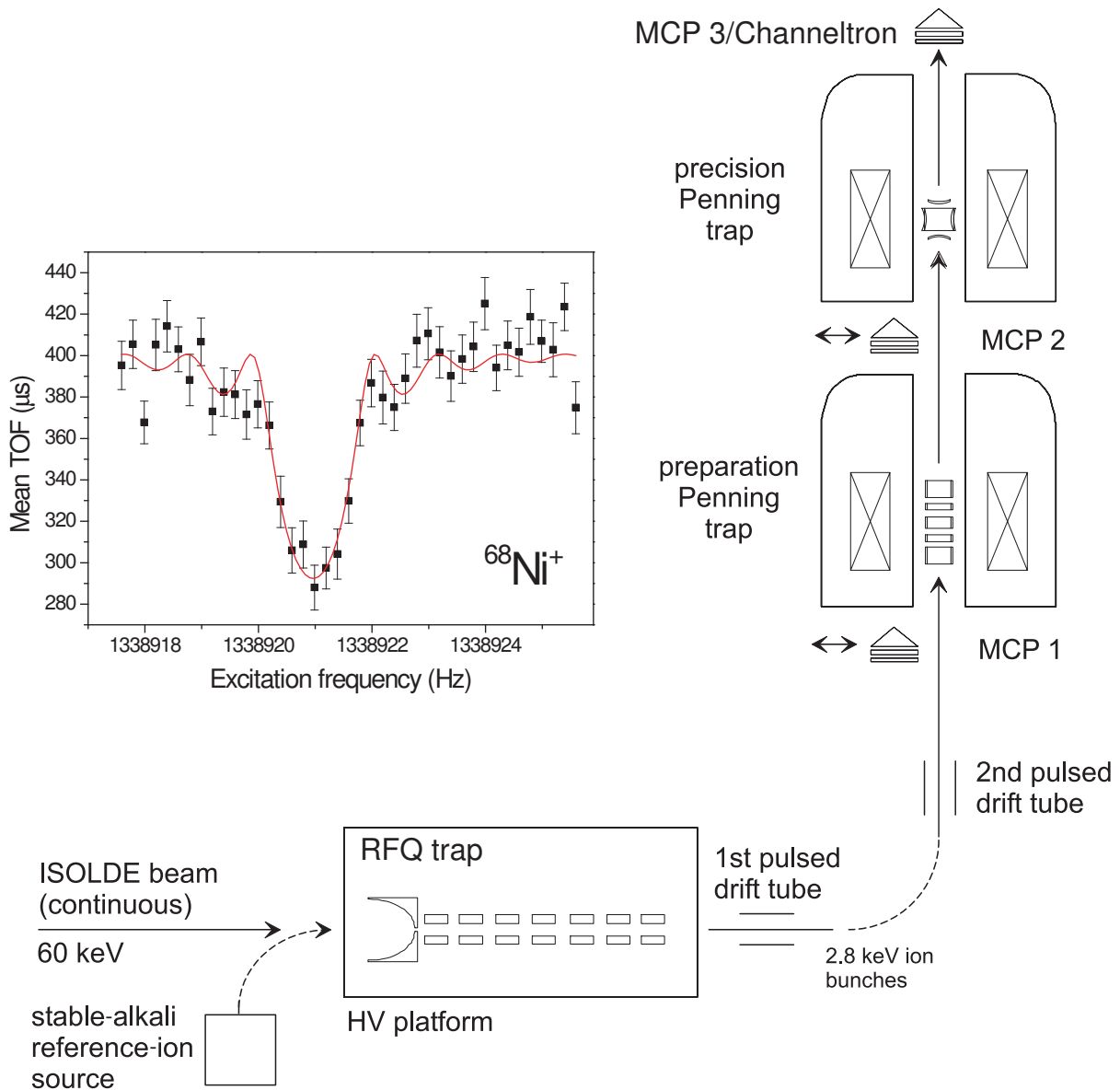}
\caption{\small{\textit{Sketch of the experimental setup of the
ISOLTRAP mass spectrometer, including the main parts: a gas-filled
linear radio-frequency quadrupole (RFQ) trap for capturing and
preparing the ISOLDE beam, a gas-filled cylindrical Penning trap
for isobaric separation, and a hyperbolic Penning trap for the
mass measurement. The micro-channel plate (MCP) detectors are used
to monitor the ion transfer and to measure the extracted-ion time
of flight (TOF) together with the channeltron detector. The inset
presents a time-of-flight (TOF) cyclotron resonance for
radioactive $^{68}$Ni$^+$ ions. }}}\label{Spe_setup}
\end{center}
\end{figure*}

ISOLTRAP is a high-precision Penning-trap mass spectrometer,
located at CERN's ISOLDE facility\,\cite{Kug00} which delivers
mass-separated beams of radionuclides.  ISOLTRAP is composed of
three main parts (see Fig.\,\ref{Spe_setup}). First, a linear
gas-filled radio-frequency quadrupole (RFQ) trap, used as cooler
and buncher, adapts the 60-keV ISOLDE ion beam to the ISOLTRAP
requirements with respect to kinetic energy, time structure, and
beam emittance\,\cite{Her01}. The second part is a gas-filled,
cylindrical Penning trap\,\cite{Rai97} in which a mass-selective
helium buffer-gas cooling technique\,\cite{Sav91} with a resolving
power of up to $10^5$ is used for isobaric cleaning. This
preparation trap is installed in a $B$=4.7\,T superconducting
magnet. Finally, the cooled ion bunch is transferred to the
precision Penning trap for isomeric separation (when required) and
mass measurement. The precision Penning trap is installed in a
second superconducting magnet ($B$\,=\,5.9\,T). The mass is
determined by measuring the true cyclotron frequency $\nu_c=qB /
(2\pi m)$ of the stored ion (see next paragraph). The magnetic
field $B$ is determined from a measurement of the cyclotron
frequency of a reference ion whose mass is well known. The setup
also includes an off-line ion source to produce stable ions, used
as reference masses.

\subsection{Mass measurement procedure}\label{isol}

Ion confinement in a Penning trap is based on the application of
an electrostatic field and a magnetic field to store ions in the
axial and radial directions, respectively. The ion motion in a
Penning trap is a superposition of three independent harmonic
oscillator modes, one in the axial direction with frequency
$\nu_z$ and two in the radial direction, $i.e.$ the cyclotron
motion with reduced frequency $\nu_+$, and the magnetron motion
with frequency $\nu_-$\,\cite{Bro86,Kon95}.  In a purely
quadrupolar electric field, the frequencies are related as
follows:
\begin{eqnarray}
&&\nu_c=\nu_++\nu_-.
\end{eqnarray}

Ion beams are alternatively delivered from ISOLDE or from an
off-line ion source and injected into the RFQ, mounted on a 60-keV
pedestal, where they are cooled and bunched. The ion bunch from
the RFQ is sent to the preparation trap. Ion collisions with the
buffer gas inside this trap first cool the axial motion. A dipolar
excitation with a frequency $\nu_-$ is then applied to increase
the magnetron radius of all ion species, making it larger than the
exit hole of the trap. To select the ions of interest, an
azimuthal quadrupole radio-frequency electric field at frequency
$\nu_c$ is applied which couples the radial modes. Since one mode
is cooled by the gas, the radius is reduced and the ion cloud is
centered.  In this way the trap works as an isobar separator with
a resolving power $R=m/\Delta m$ of $10^4$ to
$10^5$\,\cite{Rai97}.

The purified ion beam is transferred to the precision trap, where
different excitations are performed. A phase-sensitive dipolar
excitation at $\nu_-$ is applied to increase the magnetron radius
of the ion motion\,\cite{Bla03b}. If there are contaminants
(isobars or isomers), a second, mass-dependent dipolar excitation
is performed at $\nu_+$ to remove them\,\cite{Bol92}. Finally, an
azimuthal quadrupole radio-frequency field is applied to convert
the initial magnetron motion into cyclotron motion. At
$\nu_{RF}=\nu_{c}$, a full conversion is obtained, leading to an
increase of the orbital magnetic moment $\mu$ and the associated
radial kinetic energy $E=\mu B$\,\cite{Bol96}. After ejection at
low axial energy, ions pass the inhomogeneous part of the magnetic
field on their way to an MCP detector (recently replaced by a
channeltron detector\,\cite{Yaz06b}) at the top of the setup.
Since the axial acceleration in this fringe field is proportional
to $\mu \cdot\partial B/\partial z$, the shortest time of flight
(TOF) is observed for $\nu_{RF}=\nu_{c}$\,\cite{Gra80}.

The mass resolution in the precision trap depends strongly on the
conversion time used for the excitation. The line width $\Delta
\nu$ of the resonance is mainly determined by the duration of the
applied RF-field ($T_{RF}$) used to couple the two radial motions.
The relation is\,\cite{Bol96}:
\begin{equation}
\Delta\nu (FWHM) \approx \frac{0.9}{T_{RF}}.\label{FWHM}
\end{equation}
The statistical precision in the cyclotron frequency determination
is given by\,\cite{Kel03b}:
\begin{equation}\label{stat}
\frac{\delta \nu}{\nu}\propto \frac{1}{\nu T_{RF} \sqrt{N}},
\end{equation}
with $N$ being the number of ions and $R=\nu T_{RF}$ the resolving
power. With sufficiently long excitation times (few seconds), a
resolving power of up to $10^7$ can be reached. As an example of a
cyclotron frequency measurement, the inset of
Fig.\,\ref{Spe_setup} presents the time-of-flight (TOF)-resonance
curve of one of the two measurements of radioactive $^{68}$Ni. The
mean TOF of the ions as a function of the applied radio-frequency
(RF) is shown. The solid line is a fit of the well-known
line-shape\,\cite{Kon95} to the data points. This measurement was
performed with about 1000 ions using an excitation time
$T_{RF}=900$\,ms, resulting in a resolving power of $1.1\times
10^{6}$ and a relative frequency uncertainty of $\delta
\nu/\nu=6\times 10^{-8}$.

\section{Measurements of the Ni, Cu, and Ga isotopes }\label{nucli}

The nuclides $^{57,60,64-69}$Ni, $^{65-74,76}$Cu, and
$^{63-65,68-78}$Ga have been investigated with ISOLTRAP. They were
produced at ISOLDE by bombarding a uranium carbide (UC) target
with $1.4$-GeV protons from CERN's Proton Synchroton Booster. The
ionization was achieved for gallium with a tungsten (W) surface
ionization ion source and for copper and nickel with the resonance
ionization laser ion-source (RILIS)\,\cite{Kos02}. ISOLDE's
General Purpose Separator (GPS), with a mass resolving power of
about 1000 was used. The proton-rich isotopes $^{63-65}$Ga  were
measured in a different experiment using a ZrO target and ISOLDE's
High Resolution Separator (HRS), which has a mass-resolving power
of about 3000. Both targets were bombarded using pulses containing
up to $3\times 10^{13}$ protons.

The yields of nickel and copper were fairly intense at about
$10^{5}$ ions/s. The efficiency of ISOLTRAP is better than 1\% so
a beam gate was used in order to limit the number of ions sent to
the precision trap and minimize ion-ion interactions that cause
frequency shifts. The typical number of ions simultaneously stored
in the precision trap was between 1 and 8.

Despite the good yields of nickel and copper nuclides, up to three
orders of magnitude more surface-ionized gallium was present. For
the measurement of $^{68}$Ni shown in Fig.\,\ref{Spe_setup}, a
cleaning of $^{68}$Ga was applied in the preparation trap. The
ratio between the yield of $^{68}$Ga and $^{68}$Ni was ``only'' a
factor of ten which was low enough to allow an effective cleaning.
This ratio was higher farther from stability and prevented the
measurement of more neutron-rich nickel and copper since the
preparation trap was saturated by the gallium isobars. Similarly,
a significant contamination of titanium oxide prevented the
measurement of more proton-rich gallium isotopes, and the presence
of rubidium isobars made the measurement of more neutron-rich
gallium isotopes impossible.

The results from the data analysis is the ratio
$\nu_{c,ref}/\nu_c$\,\cite{Kel03a}, since the atomic mass $m$ of
the ions is calculated from the ratio between the cyclotron
frequency of the reference ion $\nu_{c,ref}$ and the cyclotron
frequency of the ion of interest $\nu_c$, the atomic mass of the
reference $^{85}$Rb\,\cite{Bra99}, and the electron mass $m_e$:
\begin{equation}
m=\frac{\nu_{c,ref}}{\nu_c}(m_{85Rb}-m_e)+m_e.
\end{equation}

All the results were evaluated in order to include them in the
Atomic-Mass Evalution (AME) table\,\cite{Wap03}. The table of
atomic masses results from an evaluation of all available
experimental data on masses, including direct measurements as well
as decay and reaction studies. The AME forms a linked network and
uses a least-squares adjustment to derive the atomic masses. Among
all nuclear ground-state properties, such an evaluation is unique
to mass measurements.

The mass values from the present measurements are presented in
Tables\,\ref{tabNi} (Ni), \ref{tabCu} (Cu), and \ref{tabGa} (Ga).
These tables give the ratio of the cyclotron frequency of the
$^{85}$Rb$^+$\,\cite{Bra99} reference mass to that of the ion of
interest. The corresponding uncertainty takes into account a
statistical uncertainty depending on the number of ions, and a
systematic error\,\cite{Kel03a}. The derived mass excess value is
indicated for comparison with the AME tables from 1995 and 2003.
Since the latest Atomic-Mass Evaluation (AME2003\,\cite{Aud03})
includes the data from this work, the influence of the ISOLTRAP
measurements is also provided. Among the 36 nuclides measured
here, the influence is 100\% for 22 of them.

\begin{center}
\begin{figure}
\includegraphics{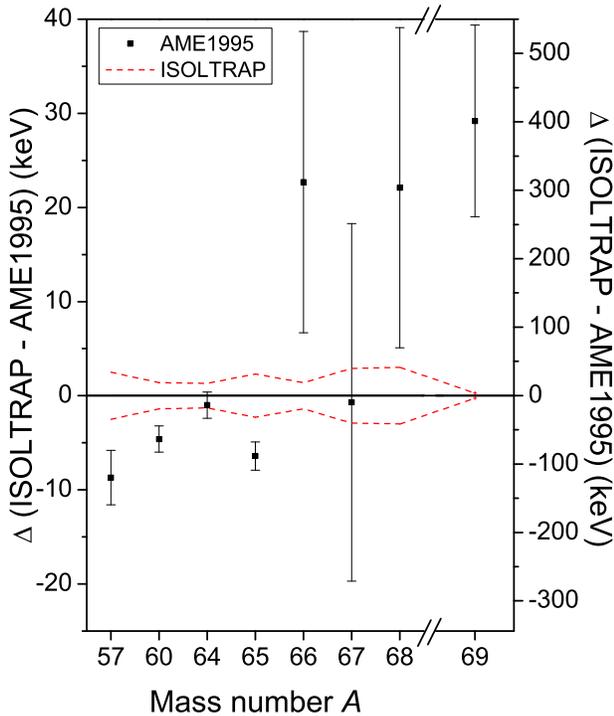}
\caption{\small{\textit{Difference between the ISOLTRAP mass
excess results for nickel isotopes and the AME1995
values\,\cite{Aud95}. Dashed lines represent the ISOLTRAP error
bars.}}} \label{fig:Ni}
\end{figure}
\end{center}

\begin{center}
\begin{figure}
\includegraphics{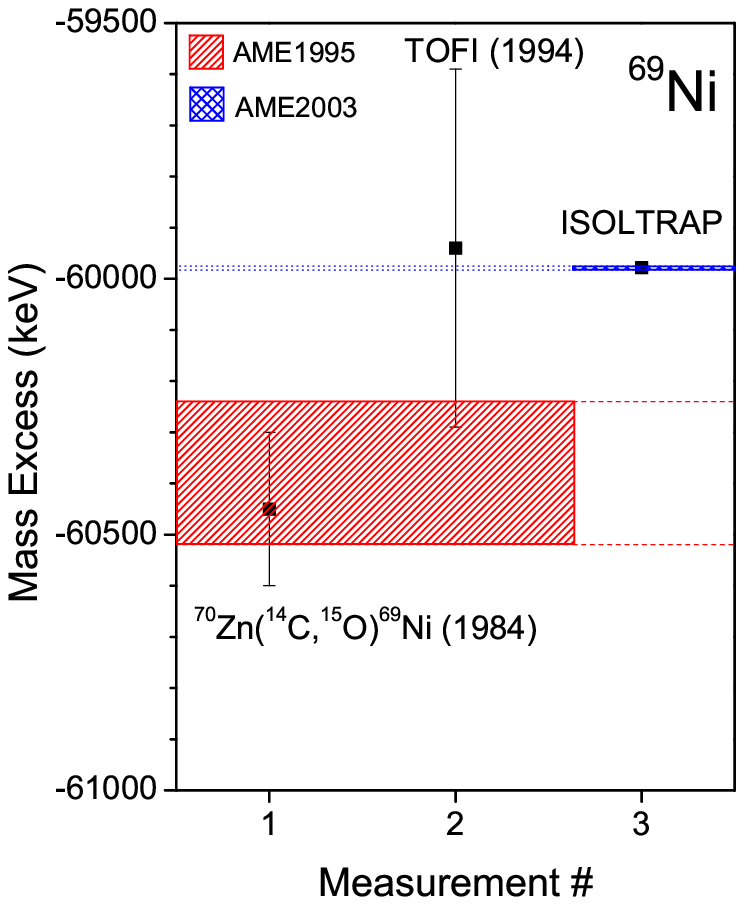}
\caption{\small{\textit{Mass excess of $^{69}$Ni determined by the
reaction $^{70}$Zn($^{14}$C,$^{15}O$)$^{69}$Ni\,\cite{Des84}, and
a time-of-flight measurement\,\cite{Sei94}, the resulting AME1995
value\,\cite{Aud95}, and the ISOLTRAP value. The AME2003
value\,\cite{Aud03} differs by 400 keV with an uncertainty 30
times smaller than the AME1995 value.}}} \label{fig:BacNi}
\end{figure}
\end{center}

The nickel results are presented in Table\,\ref{tabNi} and in
Fig.\,\ref{fig:Ni}. This figure presents the difference between
the mass excess measured by ISOLTRAP and the AME1995 values. Note
that even for the stable nickel isotopes the precision of the mass
values is improved. With the exception of $^{69}$Ni (see below)
the results are in good agreement with the 1995 table but much
more precise. The masses of $^{57,60,65}$Ni  agree with the 1995
table within the error bars, and were measured with the same order
of uncertainty. The combination of the previous value and the
ISOLTRAP measurement reduces the final uncertainty. The results
contributing to the $^{69}$Ni mass value are presented in
Fig.\,\ref{fig:BacNi}. This is a special case because it is in
strong disagreement with the AME1995 table\,\cite{Aud95}: a
difference of more than 400\,keV was observed. The AME1995 value
was derived from a $^{70}$Zn($^{14}$C,$^{15}$O)$^{69}$Ni
reaction\,\cite{Des84} and a time-of-flight
measurement\,\cite{Sei94}. The ISOLTRAP value disagrees with the
value from the reaction but is in agreement with the
time-of-flight measurement. Since the value of ISOLTRAP is
much more precise, the AME2003 includes only this value.\\

\begin{table*}
\begin{center}
\caption{\small{\textit{ISOLTRAP results for nickel isotopes:
nuclide; half life; frequency ratio $\nu_{c,ref}/\nu_c$ of nickel
isotope to reference nuclide $^{85}$Rb$^+$\,\cite{Bra99},
corresponding mass excess (ME); mass excess from AME1995; new mass
excess from AME2003; influence of the present result on the
AME2003 value. \label{tabNi}}}}

\begin{tabular}{ccccccc}
\hline
Isotopes   & Half life &$\nu_{c,ref}/\nu_c$ &  ISOLTRAP &AME1995& AME2003 &  Influence on   \\
& $T_{1/2}$& & ME (keV)  &ME (keV)  &  ME (keV)  & AME2003\\
 \hline\hline
  \\[-0.3cm]
$^{ 57}$Ni  &  35.6 h  & 0.6705736693 (316)  & -56084.2 (2.5)& -56075.5 (2.9)      & -56082.0 (1.8)     &52.0\%\\
$^{ 60}$Ni  &  Stable   & 0.7057986239 (183)    & -64472.7 (1.4)& -64468.1 (1.4)    & -64472.1 (0.6)     &16.6\%\\
$^{ 64}$Ni  &  Stable   & 0.7528734602 (163)    & -67096.9 (1.3)& -67095.9 (1.4)    & -67099.3 (0.6)     &21.9\%\\
$^{ 65}$Ni  &  2.5 h   & 0.7646753441 (285)    & -65129.0 (2.3)& -65122.6 (1.5)    & -65126.1 (0.6)     &7.8\%\\
$^{ 66}$Ni  &  55 h   & 0.7764412560 (181)   & -66006.3 (1.4)& -66028.7 (16.0)    & -66006.3 (1.4)     &100\%\\
$^{ 67}$Ni  &  21 s     & 0.7882468785 (362)   & -63742.7 (2.9) & -63742.5 (19.1)   & -63742.7 (2.9)     &100\%\\
$^{ 68}$Ni  &  29 s     & 0.8000274080 (377)   & -63463.8 (3.0) & -63486.0 (16.5)   & -63463.8 (3.0)     &100\%\\
$^{ 69}$Ni  &  12 s   & 0.8118484759 (466)    & -59978.6 (3.7) & -60380   (140)   & -59979   (4)       &100\%\\
\hline
\end{tabular}
\end{center}
\end{table*}


\begin{center}
\begin{figure}
\includegraphics[width=8.5cm]{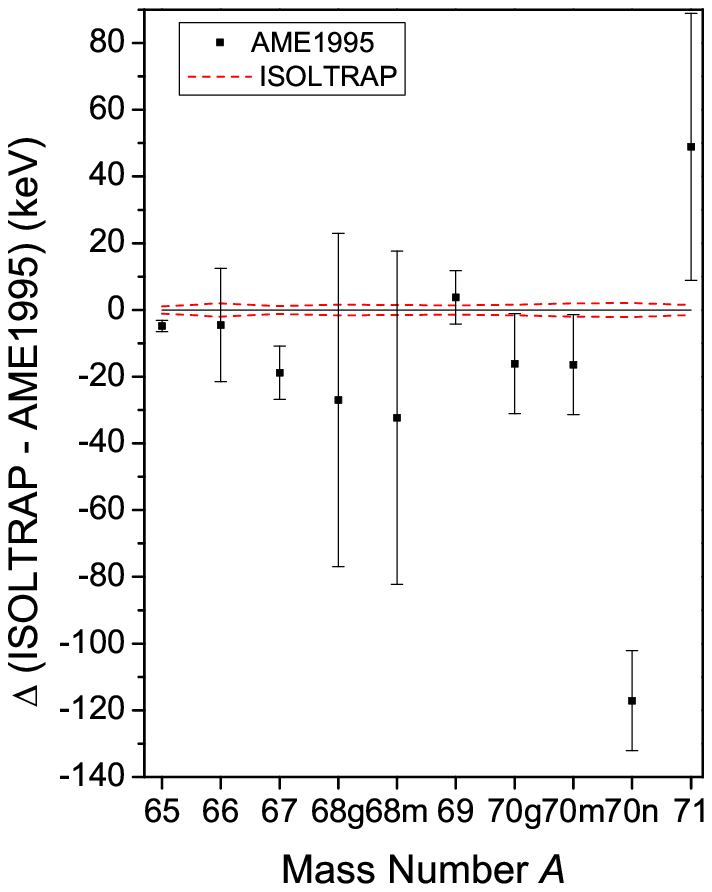}
\caption{\small{\textit{Difference between ISOLTRAP mass-excess
values for copper isotopes and the 1995 AME values\,\cite{Aud95}.
Dashed lines represent the ISOLTRAP error bars. g denotes ground
states and m,n isomeric states.}}} \label{fig:Cu}
\end{figure}
\end{center}

The copper results are listed in Table\,\ref{tabCu}, a comparison
with the AME1995 values is given in Fig.\,\ref{fig:Cu}. An
improvement of the mass uncertainty was achieved for all
investigated copper isotopes. The values are in good agreement
with previous values, except for $^{70}$Cu$^n$. This important
difference is due to an incorrect state assignment. ISOLTRAP's
high resolving power of more than $10^{6}$, in combination with
$\beta$-decay studies and selective laser ionization allowed us to
perform a clear identification of each state\,\cite{Roo04}.
Moreover, this high resolving power allowed us to resolve isomeric
states in $^{68}$Cu\,\cite{Bla04} and to measure them
independently. The masses of $^{72-74,76}$Cu were previously
unknown.
They are compared to model predictions in Section\,\ref{Models}.\\

\begin{table*}
\begin{center}
\caption{\small{\textit{ISOLTRAP results for copper isotopes:
nuclide; half life; frequency ratio $\nu_{c,ref}/\nu_c$ of copper
isotope to reference nuclide $^{85}$Rb$^+$\,\cite{Bra99},
corresponding mass excess (ME); mass excess from AME1995; new mass
excess from AME2003; influence of the present result on the
AME2003 value. Previously unknown values derived from systematic
trends are marked with \#.\label{tabCu}}}}

\begin{tabular}{ccccccc}
\hline Isotopes\footnote{g,m,n denote the ground, first excited,
and second excited state, respectively, of the nuclide.}   & Half life& $\nu_{c,ref}/\nu_c$ &  ISOLTRAP &AME1995& AME2003  & Influence   \\
 & $T_{1/2}$& & ME (keV)  &ME (keV)  & ME (keV) & on AME2003\\
 \hline\hline
  \\[-0.3cm]
$^{ 65}$Cu      &  Stable    &  0.7646483448 (139)&  -67264.5 (1.1)&-67259.7 (1.7)        & -67263.7 (0.7)    &36.8\%\\
$^{ 66}$Cu      &  5.1 m  &  0.7764380632 (257)& -66258.8 (2.0)& -66254.3 (1.7)        & -66258.3 (0.7)     &11.1\%\\
$^{ 67}$Cu      &  62 h  &  0.7882016658 (155) & -67318.8 (1.2)& -67300.2 (8.1)       & -67318.8 (1.2)     &100\%\\
$^{ 68}$Cu$^{g}$&  31.1 s   &  0.8000008176 (199) & -65567.0 (1.6)& -65541.9 (45.6)      & -65567.0 (1.6)     &100\%\\
$^{ 68}$Cu$^{m}$&  3.7 m   &  0.8000098791 (188) & -64850.3 (1.5)& -64818 (50)          & -64845.4 (1.7)     &50\%\\
$^{ 69}$Cu      &  2.8 m   &  0.8117756816 (174)& -65736.2 (1.4)& -65739.9 (8.1)        & -65736.2 (1.4)     &100\%\\
$^{ 70}$Cu$^{g}$&  45 s   &  0.8235875816 (199)& -62976.1 (1.6)& -62960.3 (14.5)       & -62976.1 (1.6)     &100\%\\
$^{ 70}$Cu$^{m}$&  33 s     &  0.8235888547 (258) & -62875.4 (2.0)& -62859 (15)          & -62875.4 (2.0)     &100\%\\
$^{ 70}$Cu$^{n}$& 6.6 s     &  0.8235906419 (272)& -62734.1 (2.1) & -62617 (15)          & -62734.1 (2.1)     &100\%\\
$^{ 71}$Cu      &  19 s   &  0.8353679363 (194)& -62711.1 (1.5)& -62764.2 (35.2)       & -62711.1 (1.5)     &100\%\\
$^{ 72}$Cu      &  6.6 s    &  0.8471819597 (182)& -59783.0 (1.4)& -60060\# (200\#)      & -59783.0 (1.4)     &100\%\\
$^{ 73}$Cu      &  4.2 s    &  0.8589690332 (491)& -58986.6 (3.9)& -59160\# (300\#)      & -58987 (4)         &100\%\\
$^{ 74}$Cu      &  1.6 s  &  0.8707837184 (779)& -56006.2 (6.2)& -55700\# (400\#)      & -56006 (6)         &100\%\\
$^{ 76}$Cu      &  640 ms   &  0.8944013229 (843)& -50976.0 (6.7)& -50310\# (600\#)      & -50976 (7)         &100\%\\
\hline
\end{tabular}
\end{center}
\end{table*}

\begin{center}
\begin{figure}
\includegraphics[width=8.5cm]{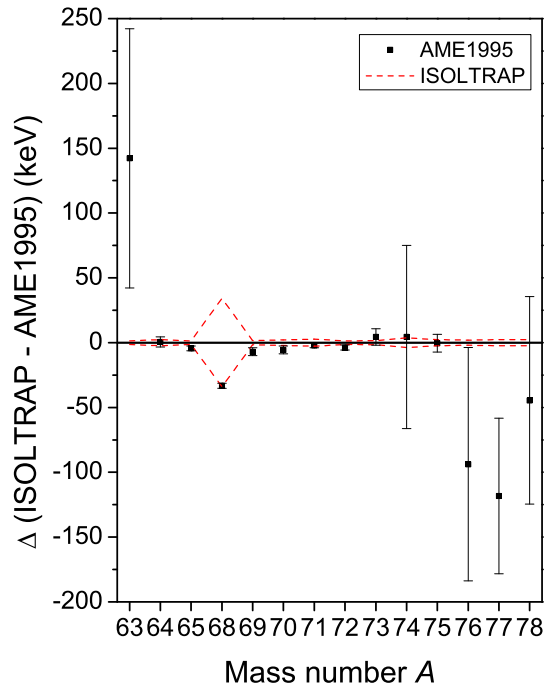}
\caption{\small{\textit{Difference between ISOLTRAP mass-excess
values for gallium isotopes and the 1995 AME values\,\cite{Aud95}.
Dashed lines represent the ISOLTRAP error bars.}}} \label{fig:Ga}
\end{figure}
\end{center}

The gallium results are presented in Table\,\ref{tabGa} and in
Fig.\,\ref{fig:Ga}. The $^{68}$Ga mass uncertainty, $\delta m/m
\approx 5.4\cdot 10^{-7}$ is much higher than for all the other
nuclides. This is due to the use of a shorter excitation time
(100~ms compared to 900\,ms for the other nuclides) and to a lack
of statistics: only 530 ions were observed, compared to at least
3000 for most of the other ones. The ISOLTRAP value is still in
agreement with the AME1995 value but has no influence. For all
other gallium isotopes measured by ISOLTRAP the uncertainty was
decreased. For five of them, it was decreased by
more than a factor of 20, and for $^{63}$Ga, almost 100 times.\\

The case of $^{74}$Ga was complicated by the possible presence of
a 9.5-second isomeric state having an excitation energy of only 60
keV (this accounts for the large AME1995 error bar in
Fig.\,\ref{fig:Ga}). Spectroscopy studies performed in parallel
with the mass measurements revealed no indication that the isomer
was produced. A two-second excitation time was used in order to
resolve this state in the precision trap but it was not seen.
Moreover, the z-class analysis\,\cite{Kel03a} was performed to
examine any dependence of the result as a function of ion number,
but revealed no indication of a contaminant. Therefore we are
confident that the present result is that of the ground-state
mass.

 \begin{table*}
 \begin{center}
\caption{\small{\textit{ISOLTRAP results for gallium isotopes:
nuclide; half life; frequency ratio $\nu_{c,ref}/\nu_c$ of gallium
isotope to reference nuclide $^{85}$Rb$^+$\,\cite{Bra99},
corresponding mass excess (ME); mass excess from AME1995; new mass
excess from AME2003; influence of the present result on the
AME2003 value.\label{tabGa}}}}

\begin{tabular}{cccccccc}
\hline
Isotopes   & Half life & $\nu_{c,ref}/\nu_c$&  ISOLTRAP&AME1995 &AME2003 &Influence   \\
 & $T_{1/2}$& &ME (keV)  &ME (keV)  & ME (keV) &  on AME2003\\
  \hline\hline
   \\[-0.3cm]
$^{ 63}$Ga  &  32 s   & 0.7412298391 (167)&  -56547.1 (1.3)&-56689.3 (100.0)       & -56547.1 (1.3)      &100\%\\
$^{ 64}$Ga  &  2.6 m  &  0.7529779275 (294)& -58834.1 (2.3)& -58834.7 (3.9)         & -58834.3 (2.0)       &75.2\%\\
$^{ 65}$Ga  &  15 m   &   0.7647065938  (176) & -62657.3 (1.4)& -62652.9 (1.8)         & -62657.2 (0.8)    &35.6\%\\
$^{ 68}$Ga  &  68 m  &  0.799981231  (431)& -67116.2 (34.1)& -67082.9 (2.0)        & -67086.1 (1.5)       &0\%\\
$^{ 69}$Ga  &  Stable   &   0.8117302720 (193)& -69327.9 (1.5)& -69320.9 (3.0)         & -69327.8 (1.2)     &65.3\%\\
$^{ 70}$Ga  &  21 m  &  0.8235125549 (272)& -68910.3 (2.2)& -68904.7 (3.1)         & -68910.1 (1.2)     &31.8\%\\
$^{ 71}$Ga  &  Stable   &  0.8352740255 (357)& -70138.9 (2.8)& -70136.8 (1.8)         & -70140.2 (1.0)      &13.3\%\\
$^{ 72}$Ga  &  14.1 h  &   0.8470706093  (182)& -68590.2 (1.4)& -68586.5 (2.0)         & -68589.4 (1.0)     &53.0\%\\
$^{ 73}$Ga  &  4.8 h   &  0.8588335898   (208)& -69699.4 (1.7)& -69703.8 (6.3)         & -69699.3 (1.7)       &100\%\\
$^{ 74}$Ga  &  8.1 m   &    0.8706314521 (469)& -68049.6 (3.7)& -68054.0 (70.7)        & -68050 (4)         &100\%\\
$^{ 75}$Ga  &  130 s    &  0.8824032092  (305)& -68464.6 (2.4)& -68464.2 (6.8)         & -68464.6 (2.4)     &100\%\\
$^{ 76}$Ga  &  33 s   &  0.8942076217  (246)& -66296.7 (2.0)& -66202.9 (90.0)        & -66296.6 (2.0)      &100\%\\
$^{ 77}$Ga  &  13 s   &   0.9059884728  (303)& -65992.4 (2.4)& -65874.1 (60.0)        & -65992.3 (2.4)     &100\%\\
$^{ 78}$Ga  &  5.1 s   &  0.9177943761  (307)& -63706.6 (2.4)& -63662.1 (80.1)        & -63706.6 (2.4)      &100\%\\
\hline
\end{tabular}
\end{center}
\end{table*}

\section{Mass-model predictions compared with new data}\label{Models}

Various models and formulae have been developed over the years to
predict properties of nuclides, particularly their mass. A review
can be found in\,\cite{Lun03} where a subset of mass models was
singled out for comparison. We have chosen to compare our
experimental data to those, as described below.

The venerable Bethe-Weizsäcker mass formula\,\cite{Wei35,Bet36},
was based on the liquid drop model and did not include shell
effects. The nuclear mass $m$ is given by

\begin{eqnarray}
m(N,Z)c^2&=&Zm_pc^2+Nm_nc^2-a_vA+a_sA^{2/3}\nonumber\\
&+&a_cZ^2A^{-1/3}+a_{sym} \frac{(Z-A/2)^2}{A},\label{eq_TH_BW}
\end{eqnarray}

where $m_p$ and $m_n$ are the proton and neutron masses, and $A$
the mass number of the nucleus. The parameters are: $a_v$ the
volume term, $a_s$ the surface term, $a_c$ the Coulomb parameter,
and $a_{sym}$ the asymmetry parameter. Note that the tabulated
masses are those of the neutral atoms, not of the bare atomic
nuclei. While inappropriate for mass predictions, it can play an
interesting diagnostic role concerning closed shell effects (see
section\,\ref{BW}).

For many years, a hybrid approach was adopted for predicting
masses based on a combination of the macroscopic liquid drop model
and microscopic (\textit{e.g.} shell) corrections.  The most
developed form of these so-called mic-mac models is the Finite
Range Droplet Model (FRDM)\,\cite{Mol95}.

The Duflo-Zuker (DZ) mass formula\,\cite{Duf99}, is a global
approach, derived from a Shell-Model Hamiltonian and gives the
best fit to the known masses. Shell-Model calculations, while
well-suited for excitation energies, are less so for mass
predictions although some efforts were made in this
direction\,\cite{Cau99}.

In the last few years, Hartree-Fock Bogolioubov (HFB) calculations
have been applied to the construction of complete mass tables.
Skyrme forces have traditionally aimed at predicting a wide range
of nuclear properties\,\cite{Flo78,Bon90,Hee95,Bon05}. The first
microscopic Skyrme-force mass formula HFBCS-1\,\cite{Ton00,Gor01}
was rapidly superceded by HFB-1\,\cite{Sam01} which, in turn, was
considerably revised, resulting in HFB-2\,\cite{Gor02}.  A
systematic study of the different adjustable parameters followed,
resulting in a series of formulas up to
HFB-9\,\cite{Sam03,Gor03,Sam04,Gor05}.

In addition to DZ and FRDM, the ISOLTRAP results are therefore
compared to HFB-2 and the recent HFB-8 (HFB-9 did not change the
mass predictions appreciably).

One characterization of a model is the root-mean-square ($rms$)
deviation from the mass values to which its parameters were
fitted, defined by

\begin{equation}
\sigma_{rms}=\frac{1}{N}\sqrt{\sum_{i=1}^{N}(m_{exp}^i-m_{th}^i)^2},
\end{equation}
where $N$ is the number of experimental $m_{exp}$ and theoretical
$m_{th}$ masses being compared. A more complete description of the
$rms$ deviation, including errors, can be found in\,\cite{Lun03}.
Table \ref{tab:rmsALL} shows $\sigma_{rms}$ for the models
compared with the AME95 table\,\cite{Aud95}, which does not
include the present ISOLTRAP results, and with
AME03\,\cite{Aud03}, which does. Our results improved the overall
agreement for the HFB models, worsened it for the Duflo-Zuker (DZ)
mass formula and for FDRM there is no change.  Examining the
isotopic chains individually, we see that in all cases the HFB
models improved and the DZ model worsened. For the FRDM, the
better fit for the gallium isotopes counters the worse fit for
copper and nickel. The differences are admittedly small (between 1
and 10\%). While it is tempting to conclude that the comparison of
the $\sigma_{rms}$ might be a demonstration of the positive
evolution of HFB-2 to HFB-8, it is important to recall that unlike
FRDM and DZ, HFB-8 was adjusted
to the masses of the AME03.\\

\begin{table}[h]
\caption{\small{\textit{The root-mean-square deviation
$\sigma_{rms}$ (in MeV) for different models: the Duflo-Zuker (DZ)
mass formula, the Finite Range Droplet Model (FRDM), and the
Hartree-Fock Bogolioubov (HFB) calculations, performed with the
AME tables of 1995 and 2003 (the latter includes the present
ISOLTRAP data). Calculations were made for the nickel, copper, and
gallium isotopes measured by ISOLTRAP. The first two rows present
the calculation for all nuclides and the following rows describe
the results for each isotopic chain separately.}}}
\label{tab:rmsALL}
\begin{center}
\begin{tabular}{cccccc}
\hline\noalign{\smallskip}
Nuclide &AME Table &DZ & FRDM & HFB-2 & HFB-8\\
 \noalign{\smallskip}\hline\noalign{\smallskip}\hline
Ni,Cu,Ga&AME95& 0.434 & 0.555 &  0.843 & 0.550\\
\hline
Ni,Cu,Ga&AME03  & 0.451 & 0.555 & 0.801 & 0.530 \\
 \noalign{\smallskip}\hline\hline
Ni & AME95&0.623 & 0.445 & 1.211& 0.732\\
\hline
Ni & AME03  & 0.640 & 0.476 & 1.174 &0.678\\
 \noalign{\smallskip}\hline\hline
Cu &AME95&0.426 & 0.471 &  0.644& 0.601\\
\hline
Cu &AME03  &0.451 & 0.530 & 0.626 &0.563\\
 \noalign{\smallskip}\hline\hline
Ga & AME95&0.280 & 0.644 &  0.654& 0.375\\
\hline
Ga & AME03  & 0.291 & 0.614 & 0.648 &0.384\\
 \noalign{\smallskip}\hline\hline
\end{tabular}
\end{center}
\end{table}

\begin{figure}
\begin{center}
\includegraphics[width=7cm]{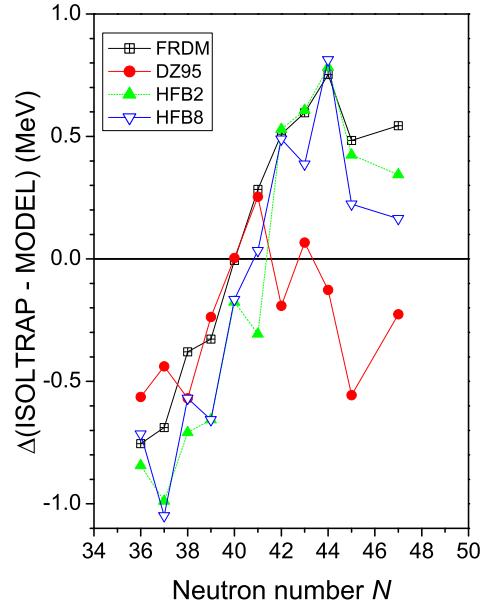}
\caption{\small{\textit{Mass difference between ISOLTRAP results
and model predictions for the copper isotopes. Note that
$^{72,73,74,76}$Cu are measured for the first time and that the
more recent parameter fit for HFB-8 included these results.}}}
\label{models_cu}
\end{center}
\end{figure}

Of particular interest for mass models is to compare predictions
as far as possible from what is already known.  In the case of the
copper isotopes presented here, four new masses were determined
and one of them ($^{76}$Cu) has five neutrons more than the most
neutron-rich previously known mass. The differences of the new
ISOLTRAP copper masses with respect to the above-mentioned models
are shown in Fig.\,\ref{models_cu}.

Despite going significantly farther from stability, it is
difficult to asses which model does a better job.  The one closest
to the new mass of $^{76}$Cu is HFB-8, however the other models
are not far away. The $rms$ errors on just the four previously
unknown masses are also similar with DZ (0.309~MeV) seeming to
follow with a better trend compared to all the others (HFB-8:
0.400~MeV; HFB-2: 0.566~MeV; FRDM: 0.603~MeV).  It is surprising
that despite all models having their parameters adjusted to the
mass tables that included those nuclides with $N<43$, those masses
are not very well reproduced locally.

%_______________________________________________________________
%%%%LIBERT text
Some nucleon-nucleon effective interactions -- like for instance
Skyrme SKM*, SLy4, or Gogny D1 -- are designed to give rise to a
realistic mean field (including pairing). They are therefore
parameterized on the ground of a few available nuclear data for
which mean field (including pairing) effects can be reasonably
disentangle from long range correlations ones (for instance,
binding energies of doubly magic nuclei only). Such approaches of
nuclei in which long range correlations are not introduced in the
mean field in an effective and somewhat uncontrolled manner do not
have as objective to give a precise mass formula at the mean (HFB)
(including pairing) level, but to constitute the mean field input
of more elaborated descriptions of nuclei considering -- at least
some -- long range correlations up to the best and therefore able
to describe ``beyond'' mean field a large class of nuclear
observable (mass formula but also low energy spectroscopy, shape
coexistence, and transitions, etc ...).
\begin{figure}
\begin{center}
\includegraphics{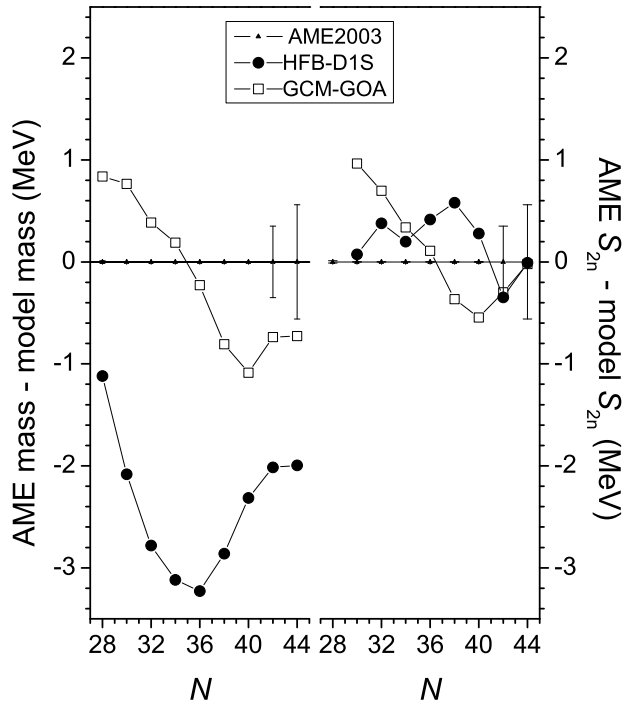}
\caption{\small{\textit{Difference of the nickel results from the
Atomic Mass Evaluation 2003 (AME2003) which already includes the
present ISOLTRAP data and those predicted by HFB-D1S (Gogny) and
GCM-GOA as a function of neutron number $N$ for (left) the mass
and (right) the two-neutron separation
energy.}}}\label{Th_two-pro}
\end{center}
\end{figure}
  In this frame, we have  performed triaxial HFB calculations, using
numerical methods and codes described in \cite{Gir83}, with the
Gogny D1S force\,\cite{Gog75,Dec80,Ber91b}.
Fig.\,\ref{Th_two-pro}\,(left) presents the differences between
the measured Ni masses and those predicted by HFB-D1S, as a
function of $N$. There is a large offset ($rms$ difference of
2.473 MeV) for the HFB-D1S masses, expected, as explained above,
specially for mid-shell nuclei where long range correlations play
an important role. Under these assumptions, we could expect at
least that the derivative of these quantities might be closer to
reality. Therefore, in Fig.\,\ref{Th_two-pro}\,(right), we have
plotted the two-neutron separation energy $S_{2n}$ [see
eq.\,(\ref{eq:s2n})] derived from the same results. The result is
encouraging, with an $rms$ deviation of only 0.508~MeV.

  In general, due to the existence of long range correlations beyond
mean field, a unique  HFB wave function is not well suited to
describe the nuclear system. Thus,  a configuration mixing
approach already described and applied with some noticeable
successes to different nuclear problems,  for instance to shape
coexistence and transitions in light mercury
isotopes\,\cite{Del94}, or  Normal-Super-deformed
phenomena\,\cite{Lib99,Del06}  has been considered. Using a
Generator Coordinate approach under Gaussian Overlap Approximation
(GCM-GOA) in a space constituted by HFB (D1S) states under axial
and triaxial quadrupole constraints allows in this model to treat
on the same footing rotation and quadrupole vibrations.
  This approach which takes explicitly into account these important
correlations, has been applied to the calculation of nickel
masses, and the results are shown in Fig.\,\ref{Th_two-pro}  for
comparison. Already the mass values (left) are greatly improved
($rms$ difference of 0.701~MeV), as are the mass derivatives
(right, $rms$ difference of 0.335~MeV). It would appear that going
beyond the mean field is to be encouraged for future mass
predictions. Works in this spirit are also underway on the ground
of Skyrme forces (see $e.g.$ \cite{Ben05}).

\section{Analysis of the mass surface around Z=29 and N=40}\label{magic}
As recalled in the introduction, Bohr and Mottelson\,\cite{Boh69}
explain that the effects of binding energy on nuclear structure
are subtle but decisive.  As such, accurate mass measurements are
important in order to finely analyse the mass surface, notably its
derivatives. In this section we examine several mass-surface
derivatives and variations.

\subsection{Study of the two-neutron separation energy}
\begin{figure*}
\begin{center}
\includegraphics[width=12cm]{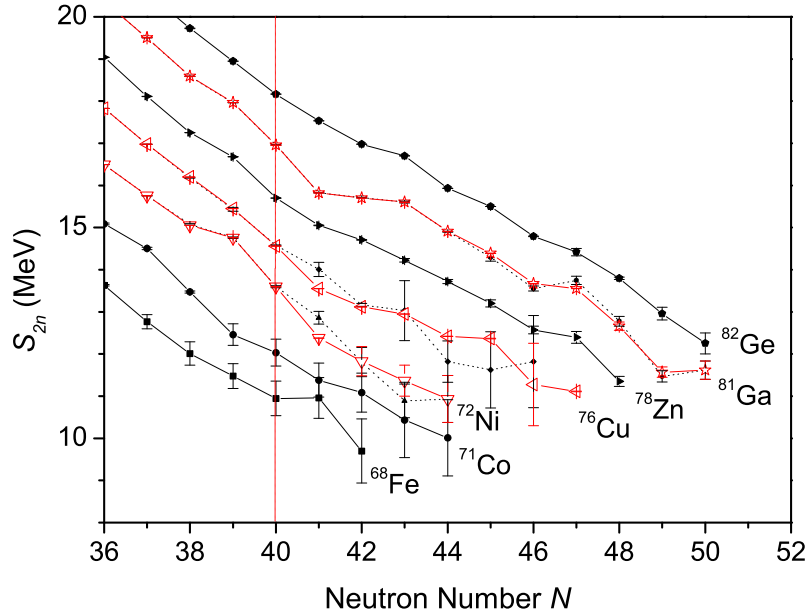}
\caption{\small{\textit{Two-neutron separation energies ($S_{2n}$)
for iron ($Z=26$) to germanium ($Z=32$) around \textit{N}\,=\,40.
Dashed lines correspond to the data before the ISOLTRAP
measurements. Points with large error bars were not directly
measured by ISOLTRAP but their value was changed by the link to
the measured masses.}}} \label{fig:S2n}
\end{center}
\end{figure*}

The two-neutron separation energy ($S_{2n}$) given by
\begin{equation}
\label{eq:s2n} S_{2n}(N,Z)= B(N,Z)-B(N-2,Z),
\end{equation}
with $B$ for the binding energy, is remarkable for its regularity
between shell closures. Generally, $S_{2n}$ decreases smoothly
with $N$ and shell effects appear as discontinuities. In the past,
discontinuities of $S_{2n}$ versus \textit{N} were often traced to
inaccurate $Q_{\beta}$ endpoint measurements and measurements with
more reliable, direct techniques restored the regularity (see, for
example,\,\cite{SCS01} for the area around $^{208}$Pb). Hence,
part of the motivation was to confirm any mass surface
irregularities in the $N=40$ region. Fig.\,\ref{fig:S2n} presents
the $S_{2n}$ values, from $N=36$ to 50, prior and after the
ISOLTRAP mass measurements. Most of the irregularities
\textit{e.g.} at $N=41$ for gallium are confirmed. Moreover, the
plot reveals a deviation from the linear trend between $N=39$ and
$N=41$ for nickel, copper, and gallium. Also irregularities for
gallium ($N = 46 - 49$) and copper ($N = 43 - 46$) are visible.

\begin{figure}[h]
\begin{center}
\includegraphics{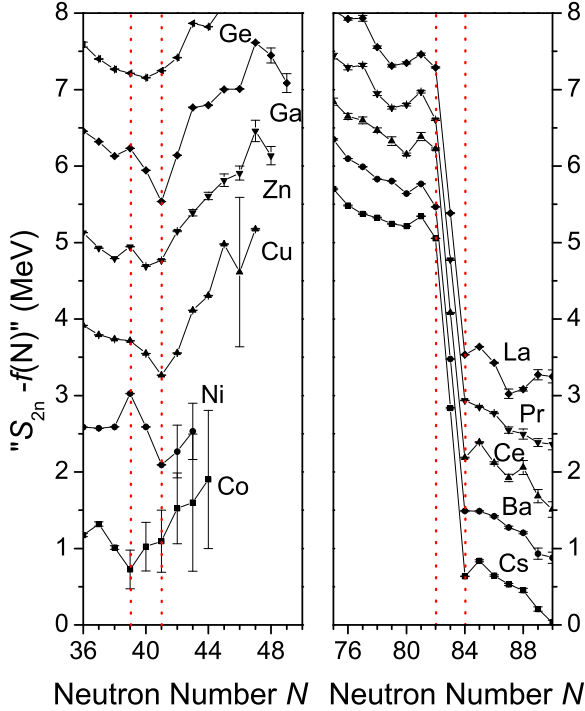}
\caption{\small{\textit{Two-neutron separation energies ($S_{2n}$)
minus a linear function of \textit{N} around $N=40$ (left), and
the strong shell closure $N=82$ (right), for comparison.}}}
\label{fig:S2n-N}
\end{center}
\end{figure}

To study the structure more closely we subtract a linear function
of \textit{N} determined by the $S_{2n}$ slope preceding the
purported shell closure. The resulting reduced $S_{2n}$ values are
presented in Fig.\,\ref{fig:S2n-N} in the region of $N=82$ (for
comparison) and $N=40$. The $N=82$ shell closure is clearly
visible on this plot: there is a change of slope between $N=82$
and $N=84$. From these observations we can analyse the behavior in
the $N=40$ region: there is a similar effect between $N=39$ and
$N=41$ where the break can be seen at $N=39$ and not at $N=40$,
surprising for an odd number. The magnitude of this decrease is
far smaller (between 500\,keV and 1\,MeV) than the one for the
major shell closure at $N=82$ (around 4\,MeV). A similar structure
is seen between $N=39$ and $N=41$ for nickel, copper, and gallium,
but this is not an indication of shell closure.  It is strange
that the same structure is visible for both nickel (even $Z$) and
gallium (odd $Z$) whereas germanium is smooth and little is seen
in the case of zinc.  Further measurements to reduce the
uncertainty on the neighboring cobalt isotopes will be needed.

%%%%%%%%%%%%%%%%%%%%%%%%%%%%%%%%%%%%%%%%%%%%%%%%%%%%%%%%%%%%%%%%%%%%%%%%%%%%%%%%%%%%%%%
%--------------------------------Shell gap---------------------------------------------
%%%%%%%%%%%%%%%%%%%%%%%%%%%%%%%%%%%%%%%%%%%%%%%%%%%%%%%%%%%%%%%%%%%%%%%%%%%%%%%%%%%%%%%

\subsection{The shell gap}
\begin{figure}
\begin{center}
\includegraphics{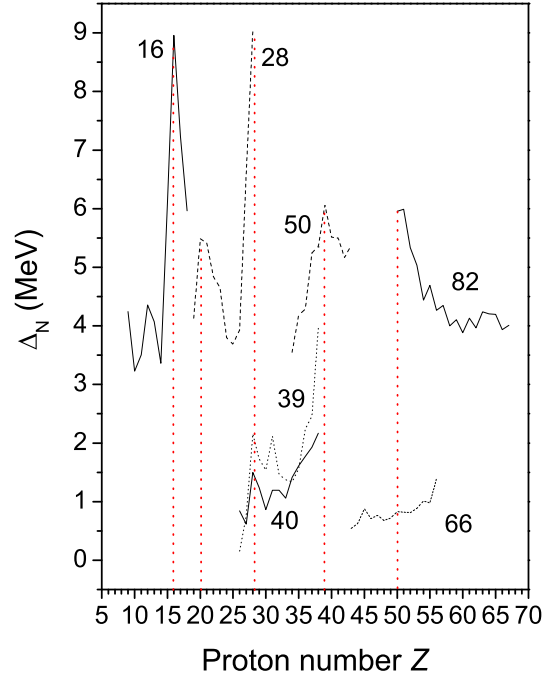}
\caption{\small{\textit{(a) Shell gap as a function of the proton
number \textit{Z} for different magic and mid-shell neutron
numbers \textit{N}. N=16, 28, 50, 82 correspond to shell closures,
$N=39$ and $66$ are exactly between two shell closures (called
mid-shell), $N=40$ is under investigation.  Data are from
\cite{Aud03}.}}} \label{fig:SG_All}
\end{center}
\end{figure}

The neutron shell gap, defined as
\begin{eqnarray} \label{eq}
\Delta_{N}(N,Z)&=&S_{2n}(N,Z)-S_{2n}(N+2,Z)\\
&=&2B(N,Z)-B(N-2,Z)-B(N+2,Z),\nonumber
\end{eqnarray}
is a good indicator of shell strength. The shell gap definition is
usually only valid for spherical nuclides, \textit{i.e.} around
magic numbers. Here, we examine the case of $N=40$ and also
investigate how mid-shell gaps compare in strength and
comportment. Fig.\,\ref{fig:SG_All}, calculated from AME2003
data\,\cite{Aud03}, shows the shell gap as a function of the
proton number $Z$ for for various $N$. This highlights the large
shell gap values for magic neutron number with peaks at magic
\textit{Z}. It also shows that for $N=50$ there is a peak at
$Z=39$, and not $Z=40$, which is known to be semi-magic. This
behavior is probably due to the odd-even effect in the two-proton
separation energy $S_{2p}$. Not surprisingly, the mid-shell-gap
($N=39$, 66) energies are quite small.  From this point of view,
the case of $N=40$ resembles a mid-shell rather than a magic
number.

\begin{figure}[p]
\begin{center}
\includegraphics[width=8.5cm]{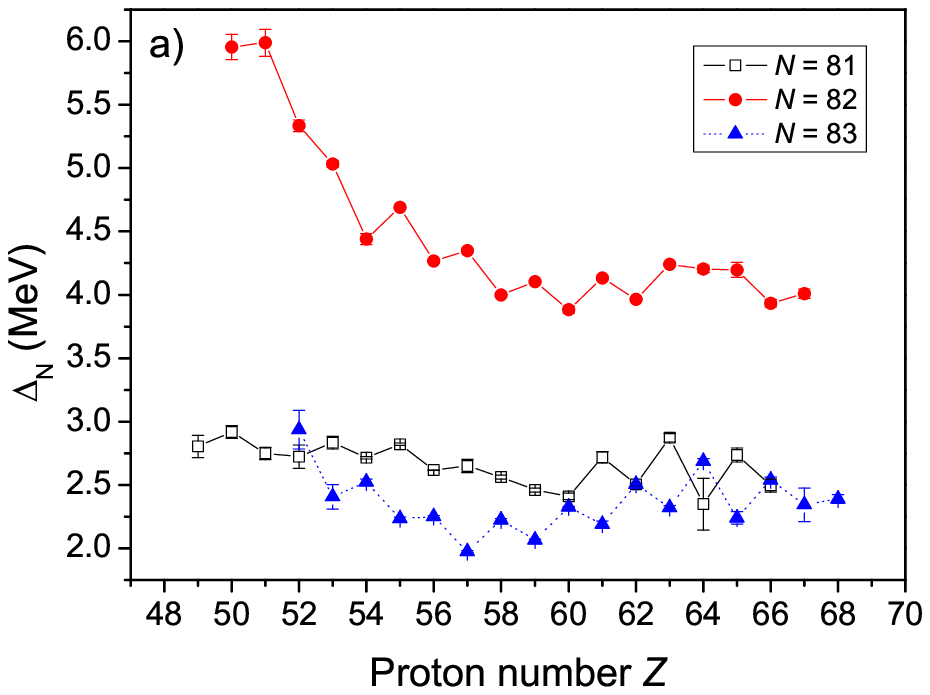}
\includegraphics[width=8.5cm]{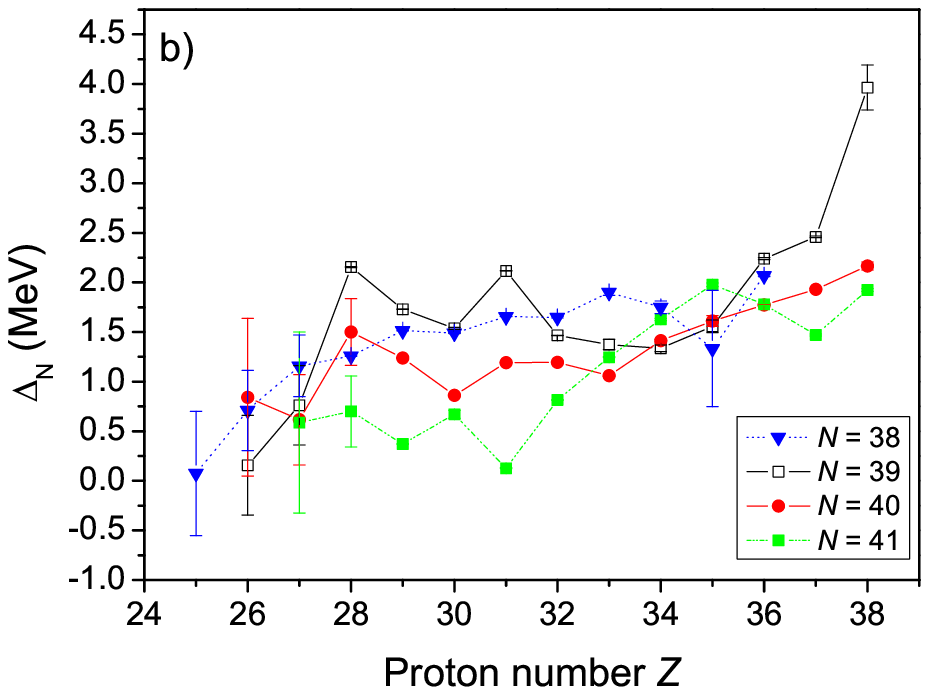}
\includegraphics[width=8.5cm]{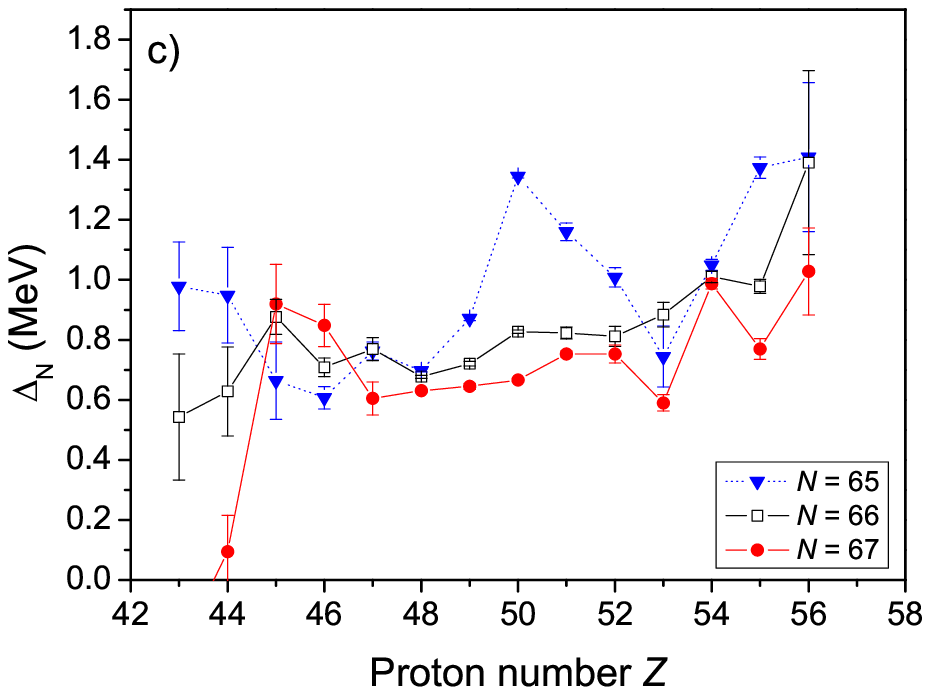}
\caption{\small{\textit{Shell gap as a function of the proton
number $Z$ for a) $N=81-83$ with the N=82 magic number well
distinguished from N=81 and 83, b) $N=38-41$, and c) $N=65-67$
with N=66 representing a mid-shell number in between N=50 and
82.}}} \label{fig:SG82}
\end{center}
\end{figure}

Fig.\,\ref{fig:SG82} shows the details of adjacent shell gaps
$\Delta_N$ as a function of the proton number $Z$ for different
regions: (a) around a shell closure, (b) in the region of
interest, and (c) in a mid-shell region. In
Fig.\,\ref{fig:SG82}(a), the behavior of a strong shell closure is
shown for $N=82$ which is a magic number: there is a large
difference between $N=82$ and $N=81$, 83 and the corresponding
enhanced shell gap for the case of magic $Z=50$.
Fig.\,\ref{fig:SG82}(c) shows the behavior of the mid-shell region
around $N=66$ (exactly in between two shell closures: 50 and 82):
the neutron shell gap for $N=66$ is between the one for $N=65$ and
$N=67$. Fig.\,\ref{fig:SG82}(b) presents the shell gap around
$N=40$. For $N=40$ a strong difference (like for $N=82$) is not
visible and \textit{N}\,=\,40 is distinct from neither $N=39$ nor
41. Note that the $N=39$ mid-shell gap is larger than those of
$N=38$ and 40 for several values of \textit{Z}, especially for
$Z=28$, unlike the $N=66$ mid-shell behavior. This shows that
$N=38$, 39, and 40 do not have the behavior we could have expected
from observation in other mass regions. However, in summary, no
shell closure at $N=40$ is observed.

%%%%%%%%%%%%%%%%%%%%%%%%%%%%%%%%%%%%%%%%%%%%%%%%%%%%%%%%%%%%%%%%%%%%%%%%%%%%%%%%%%%%%%%
%---------------------------------pairing energy----------------------------------------
%%%%%%%%%%%%%%%%%%%%%%%%%%%%%%%%%%%%%%%%%%%%%%%%%%%%%%%%%%%%%%%%%%%%%%%%%%%%%%%%%%%%%%%

\subsection{The pairing gap}
The pairing gap from the four-point formula \cite{Bar57}
$\Delta_{4}(N,Z)$
\begin{eqnarray}
\Delta_{4} (N,Z)= \frac{(-1)^{N}}{4}
\Big(&{}&B(N+1)-3B(N)\nonumber\\
&+&3B(N-1)-B(N-2)\Big) \label{eq_D4}
\end{eqnarray}
was chosen to study the pairing-energy behavior. A peak is
expected for magic numbers and a trough at mid-shell.

\begin{figure}
\begin{center}
\includegraphics{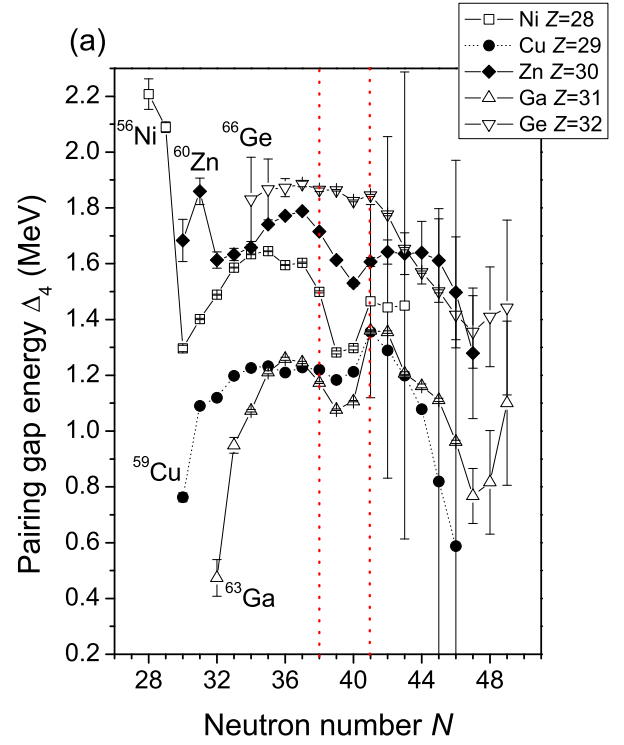}
\includegraphics{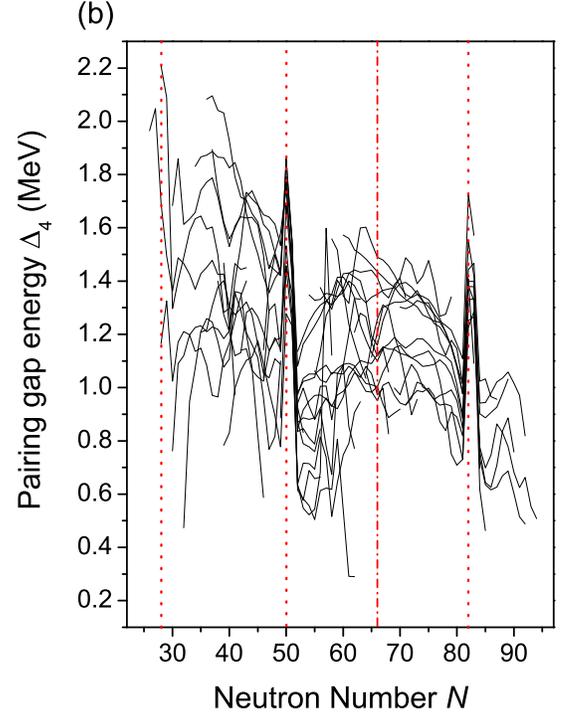}
\caption{\small{\textit{(a) Pairing gap energy as a function of
neutron number for the investigated elements: nickel, copper, and
gallium, as well as zinc and germanium. (b) Pairing gap energy as
a function of neutron number for Z=27-59. Shell closures at N=28,
50, and 82 are clearly visible, the N=66 mid-shell is
indicated.}}} \label{fig:pair40}
\end{center}
\end{figure}
The pairing gap as a function of neutron number is presented in
Fig.\,\ref{fig:pair40}(a) for $Z=28-32$.  At the $N=39$ mid-shell,
there is a trough for \textit{Z}\,=\,31 - but not for
\textit{Z}\,=\,29. A similar behavior is seen at $N=66$ (82-50
mid-shell).  The odd-$Z$ nuclides have a lower pairing gap and
while germanium ($Z=32$) shows no particular structure, nickel
($Z=28$) shows a strong mid-shell trough and not a peak that would
indicate a shell closure, as shown in Fig.\,\ref{fig:pair40}(b)
where shell closure at $N=28, 50,$ and 82 are clearly visible.

%%%%%%%%%%%%%%%%%%%%%%%%%%%%%%%%%%%%%%%%%%%%%%%%%%%%%%%%%%%%%%%%%%%%%%%%%%%%%%%%%%%%%%%
%---------------------------------B-W formula----------------------------------------
\subsection{Comparison with the Bethe-Weizs\"acker formula}\label{BW}

The Bethe-Weizs\"acker formula was given in eq.\,(\ref{eq_TH_BW}).
We adapt the version of Pearson\,\cite{Pea01}, with a pairing term
of Fletcher\,\cite{Fle03}. Thus, the binding energy per nucleon is
given by
\begin{eqnarray} \label{eqBW}
\frac{E_{nuc}}{A}&=&a_{vol}+a_{sf}A^{-1/3}+\frac{3e^{2}}{5r_{0}}Z^{2}A^{-4/3}\nonumber\\
&{}& +(a_{sym}+a_{ss}A^{-1/3})I^{2}\nonumber\\
&{}&+a_{p}A^{-y-1}\Big(\frac{(-1)^{Z}+(-1)^{N}}{2}\Big),
\end{eqnarray}
with $I=(N-Z)/A$. The parameters are $a_{vol}\,=\,-15.65$\,MeV,
$a_{sf}\,=\,17.63$\,MeV, $a_{ss}\,=\,-25.60$\,MeV which is the
parameter of surface symmetry term introduced by Myers and
Swiatecki\,\cite{Mye66}, $a_{sym}\,=\,27.72$\,MeV,
$r_{0}\,=\,1.233$\,fm with $r_0$ the constant used in the radius
estimation $R\approx r_0A^{1/3}$, $a_{p}=-7$ MeV the pairing term,
and $y=0.4$. This formula contains no specific term for shell
effects so the formula may not be a good way to predict exotic
mass values. However this makes it a ``neutral'' indicator for
shell structures (see\,\cite{Gue05b}).

To this end, the modified Weizs\"acker formula [eq. (\ref{eqBW})]
is subtracted from known masses (divided by $A$). The difference
between the experimental values and the formula clearly reveals
the shell closures at $N=28$, 50, 82 and 126, reaching up to 15
MeV for $N=50$ and $N=82$ (see Fig.~1 in\,\cite{Pea01}).

\begin{figure}
\begin{center}
\includegraphics{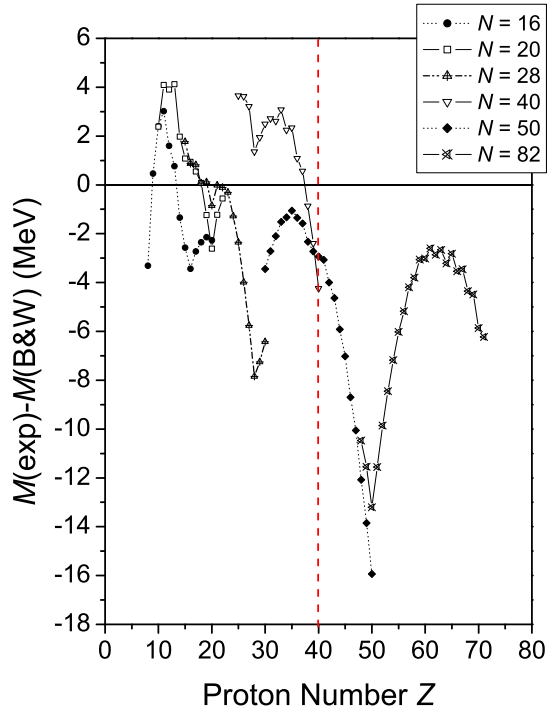}
\caption{\small{\textit{Difference between the experimental mass
values from this work and from AME2003 data\,\cite{Aud03} and
theoretical masses from the ``Bethe-Weizs\"acker formula'' as a
function of proton number, for several magic neutron numbers and
for N=40.}}} \label{fig:Z}
\end{center}
\end{figure}

Fig.\,\ref{fig:Z} presents the difference between the experimental
results obtained from this work (complemented with AME2003 data)
with the ``Bethe-Weizs\"acker formula'' [eq.\,(\ref{eqBW})] as a
function of $Z$ for various magic neutron numbers, including
$N=40$. As with the shell gaps, the cases where $N=Z$ show the
strongest effects, as does the case of $^{132}_{50}$Sn$_{82}$.
Interestingly enough, the case of $^{68}_{28}$Ni$_{40}$ does show
a dip of about 2~MeV, although only about 20\% the effect of
$^{132}_{50}$Sn$_{82}$.

\begin{figure}
\begin{center}
\includegraphics{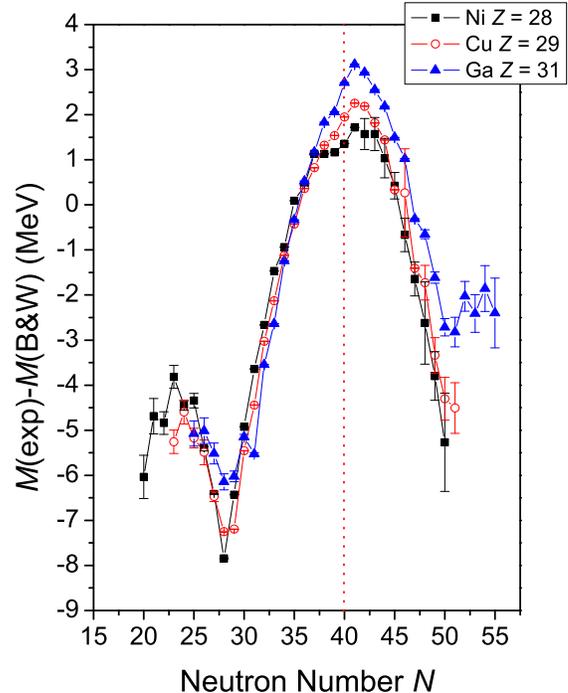}
\caption{\small{\textit{Difference between the masses predicted by
the Bethe-Weizs\"acker formula (eq. \ref{eqBW}) and the
experimental values as a function of $N$ for $Z=28$, 29, and 31.
Data are from this work complemented by \cite{Aud03}.}}}
\label{fig:BW40}
\end{center}
\end{figure}

When the difference in mass values is examined isotopically as a
function of neutron number (Fig.\,\ref{fig:BW40}), however, there
is no indication of a shell, or even sub-shell closure.  The
pseudo-parabolic behavior of the curve in Fig.\,\ref{fig:BW40}
shows some indentation around $N=40$ but nothing that we could
claim to be ``magic''.

\section{Conclusion}
The high-precision mass measurements performed at ISOLTRAP on over
30 short-lived neutron-rich isotopes of nickel, copper, and
gallium have allowed us to rather finely study the mass surface --
and its derivatives -- around the interesting region of $Z=28$ and
$N=40$. No behavior resembling that of known magic numbers has
been found, unlike the analog case of $Z=40$, where the $N=56$
sub-shell closure is visible. As much as an $N=40$ ($d_{5/2}$)
sub-shell could exist, there is no clear indication for such a
sub-shell closure from these measurements.  While the pairing gap
energy clearly indicates that there is no shell closure in this
region, a competing mid-shell stabilization effect might be
present. The comparison with the Bethe-Weizs\"acker formula shows
some fine structure around $N=39,40$ but no indication of the
presence of a shell, or sub-shell closure. The shell gap
evaluation shows anomalous behavior for $N=39$ as well as for
$N=40$, perhaps due again to the competition between a sub-shell
closure at 40 and the mid-shell at 39.

Recalling again the words of Bohr and Mottelson, ``it is
relatively difficult to discern the nuclear shell structure as
long as the main information on nuclei is confined to binding
energies''. While they are a necessary ingredient, it is not
sufficient for explaining the problem at hand since the binding
energies are in opposition with results on the
$B(E2)$\,\cite{Sor02}. Thus, more detailed spectroscopy
measurements, including the $g-$factor, as suggested by Langanke
\textit{et al.}\,\cite{Lan03}, and more theoretical work, are
called for to understand the various phenomena arising from
mass-surface studies.

\begin{acknowledgments}
The authors thank the ISOLDE technical group for their assistance
and acknowledge the GDR {\it Noyaux Exotiques}.  This work was
supported by the German Federal Ministry for Education and
Research (BMBF) under contract no. 06GF151, by the Helmholtz
Association of National Research Centers (HGF) under contract no.
VH-NG-037, by the European Commission under contract no.
HPRI-CT-2001-50034 (NIPNET), by the ISOLDE Collaboration, by the
Marie Curie fellowship network HPMT-CT-2000-00197, and by the
French IN2P3.
\end{acknowledgments}

\end{document}